\documentclass[12pts]{elsart3p}
\usepackage{graphicx}
\usepackage{mathpazo}
\newcommand{\comment}[1]{}
\usepackage{amssymb}  
\usepackage{color} 
\begin{document}
\begin{frontmatter}
\title{Position reconstruction of acoustic sources with the AMADEUS Detector}
\author{C. Richardt}
\ead{carsten.richardt@physik.uni-erlangen.de}
\author{G. Anton}
\author{K. Graf}
\author{J. H\"o\ss l}
\author{U. Katz}
\author{R. Lahmann}
\author{M. Neff}
\address{Universit\"at Erlangen
ECAP (Erlangen Centre for Astroparticle Physics),
Erwin Rommel Str. 1,
91058 Erlangen, 
Germany}

\begin{abstract}
This article focuses on techniques for position reconstruction of acoustic point sources with the 
AMADEUS setup consisting of 36 acoustic sensors in the Mediterranean Sea.
The direction reconstruction of an acoustic point source utilizes the information of the 6 small-volume hydrophone clusters of 
AMADEUS individually.  Source position reconstruction is then done by combining the directional information of each cluster.
The algorithms for direction and position reconstruction are explained and demonstrated using data taken in the deep sea.
\end{abstract}

\end{frontmatter}

\section{Introduction}

\begin{figure}[ht]
\begin{center}
  \includegraphics[angle=0,width=0.49\textwidth]{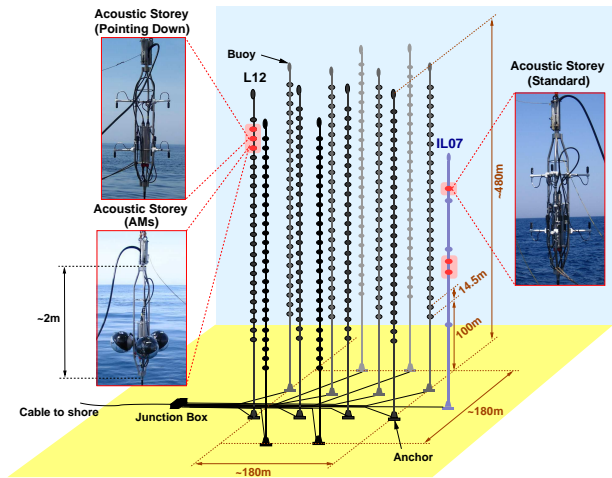}
  \caption {\label{fig:antares} { The ANTARES detector.  The AMADEUS clusters are located on the 
 Instrumentation Line (IL07) and Line 12 (L12). }}
\end{center}
\end{figure}

AMADEUS \cite{arena06} (Antares Modules for Acoustic Detection Under the Sea) is a system for acoustic detection 
within the ANTARES \cite{antares_proposal} (Astronomy with a 
Neutrino Telescope and Abyss environmental RESsearch) infrastructure. The former is designed to investigate the feasibility of 
acoustic detection of ultra high energy neutrinos. The ANTARES experiment is located $25\,$km off the southern French coast at a water depth of $2500\,$m 
and consists of $12$ vertical support structures, so called lines, of $480\,$m length.  They are anchored 
to the sea floor over an area of about $(180\times 180)\,$m$^2$, held vertically by buoys and support a total of $\sim 900$
photo-multipliers. An additional line, the Instrumentation Line, is used for environmental monitoring and 
multidisciplinary research. The AMADEUS system consists of six local clusters of six sensors each.  Three of these clusters 
are located on the 12th line while the other three clusters are located on the Instrumentation Line (IL07), see Fig. \ref{fig:antares}. 
AMADEUS was designed to investigate short and long distance correlations, study the deep sea noise, signal filtering 
and source positions reconstruction techniques, all necessary to evaluate the feasibility of acoustic particle detection with 
a possible future large volume detector. Using local clusters allows for correlation studies 
ranging from about $1\,$m to $350\,$m, and also offers the advantage of quick direction reconstruction, 
thereby simplifying source position reconstruction.

\section{Direction reconstruction}
\label{sec:dir}
Each of the six local clusters of AMADEUS comprises six hydrophones.  The hydrophones within a local cluster have typical distances of $1\,$m 
between the sensors 
and thus a transit time of $<1\,$ms for an acoustic wave front.  This allows for quick direction reconstruction of a passing pressure wave.  
Two methods for direction reconstruction with the individual clusters will be discussed. One of the methods is beam\-forming and the other one 
uses the time differences of the detected signals in the individual acoustic sensors of a storey. 

\subsection{Beamforming}
\label{sec:beam}

\begin{figure}[ht]
\begin{center}
  \includegraphics[angle=0,width=0.49\textwidth]{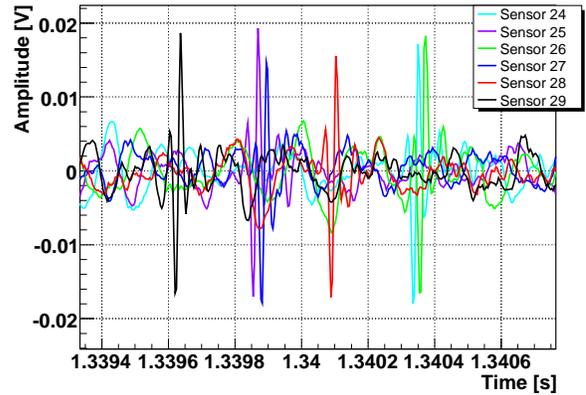}
  \caption {\label{fig:sig} {Signals recorded by one local cluster, consisting of six hydrophones.  }}
\end{center}
\end{figure}

\begin{figure}[ht]
\begin{center}
  \includegraphics[angle=0,width=0.49\textwidth]{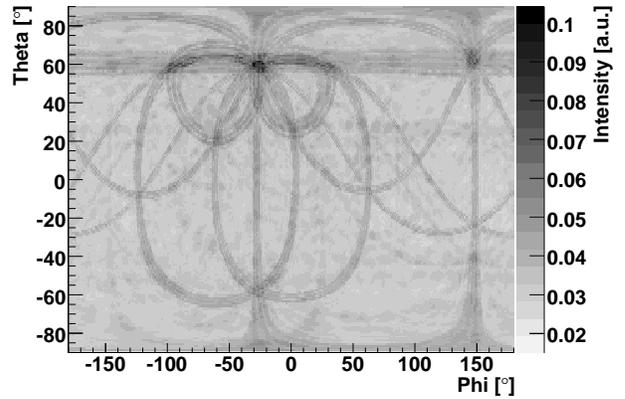}
  \caption {\label{fig:beam} {Output of the beam-forming algorithm applied to the signal shown in Fig. \ref{fig:sig} 
	The maximum, indicating the signal direction, can be identified at $\theta = 59^\circ$ and  $\varphi=-29^\circ$.}}
\end{center}
\end{figure}

This method requires the knowledge of the hydrophone positions, within a cluster, as well as synchronized data.
Beamforming is done by creating a sound intensity plot scanning all directions in space ($4\pi$) for the actual source direction.  
Given the six hydrophone coordinates $\vec{r_n}\mbox{  } (n=1,2...,6)$, the signal $p_n$ of every hydrophone 
will be shifted in time corresponding to the difference in path length of the sound wave reaching the respective hydrophone.
Hence every direction in space corresponds to a set of time differences $\Delta t_n$ in the data.  
For a direction $\vec{k}$, the beam-forming output at time $t$ is given by
\begin{equation}
\label{eq:beamer}
	b(\vec{k},t)=\sum_{n=1}^N w_np_n(t-\Delta t_n(\vec{k}))\mbox{ , }
\end{equation}
where $w_n$ represent weighting factors for every individual hydrophone.  These factors are adjusted to match 
the directional sensitivity of each hydrophones. For the following calculations $w_n\equiv 1$ was used.
The time differences $\Delta t_n$ are computed assuming a plane wave\footnote{The assumption of a plane wave is a sufficiently good 
approximation if the distances are large compared to the dimensions of the hydrophone antenna.}.
The algorithm scans $4\pi$ with a predefined step size in $\theta, \varphi$ by applying the calculated time differences to the data, assuming a constant speed of sound. 
Once all directions are scanned, the maximum value of the produced output identifies the direction of the signal. 
For this direction all signals are shifted such that they add up constructively.
The application of the algorithm increases the signal to noise ratio for the direction the signal comes from, creating the mentioned maximum. 
Figure \ref{fig:beam} shows the output of the beamforming algorithm applied to the signal in Fig. \ref{fig:sig}. 

The beam-forming output (Fig. \ref{fig:beam}) shows a well defined maximum at $\theta = 59^{\circ}$ and $\varphi = -29^{\circ}$. 
The error is governed by the binning of the algorithm and not by the algorithm itself.
Local maxima and the visible patterns in Fig. \ref{fig:beam} correspond to directions where the time 
shift results in the constructive addition of two or more signals.  

The advantage of this method is that it can be applied to untriggered data. A disadvantage of this method is that it is rather 
time consuming.

\subsection{Time difference method}
\label{sec:tdiff}

Another method for directional reconstruction is using the time differences between signals, detected by the hydrophones in a cluster, of a passing pressure wave.  
This method requires, in addition to the knowledge of the hydrophone position and synchronized data, a signal identification.
Once a signal is identified, e.g. by a simple threshold trigger for large signal amplitudes, the signal arrival times are used for the direction reconstruction.
Direction reconstruction is done by comparing the measured signal times to expected signal arrival times.
The expected time $t_{n_{expected}}(\theta,\varphi)$ for an arriving wave front is taken from a lookup table and subtracted from the measured time. 
That is done for $4\pi$ in the desired angular resolution, where the minimum indicates the reconstructed direction:

\begin{equation}
\label{eq:beamer}
	min\left\{ \tau^2:=\sum_{n=1}^6 \left( t_{n_{measured}}-t_{n_{expected}}(\theta,\varphi)\right)^2 \right\}\mbox{.}
\end{equation}

\begin{figure}[ht]
\begin{center}
  \includegraphics[angle=0,width=0.49\textwidth]{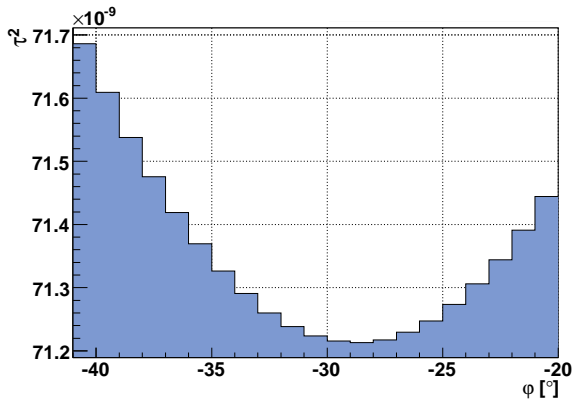}
  \caption {\label{fig:chiX} {Cut through the $\theta - \varphi$ plane at $\theta_{min}$ of the minimum computed with the time difference method.}}
\end{center}
\end{figure}

\begin{figure}[ht]
\begin{center}
  \includegraphics[angle=0,width=0.49\textwidth]{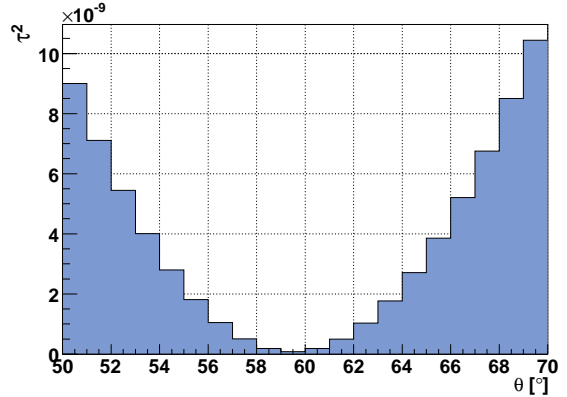}
  \caption {\label{fig:chiY} {Cut through the $\theta - \varphi$ plane at $\varphi_{min}$ of the minimum computed with the time difference method. }}
\end{center}
\end{figure}

This technique is similar to beam-forming, but much faster. Applied to the data in Fig. \ref{fig:sig} the algorithm 
produces the $\tau^2$ data shown in Fig. \ref{fig:chiX} and \ref{fig:chiY}. A clear minimum can be identified in both $\theta$ and $\varphi$.
The values match the ones obtained by the beamforming algorithm. The error is again governed by the binning of the algorithm.

Though the time difference method is much faster, beam-forming has the advantage of not having to identify a signal.

\begin{figure}[ht]
\begin{center}
  \includegraphics[angle=0,width=0.49\textwidth]{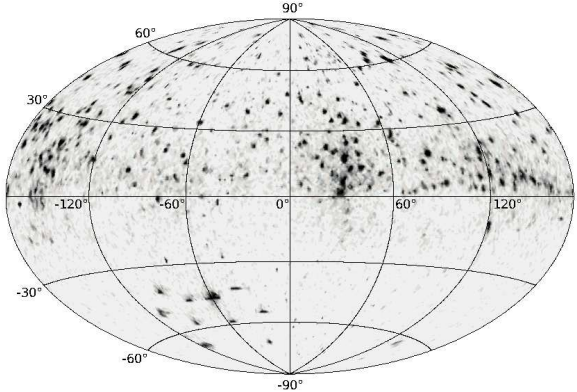}
  \caption {\label{fig:sky} {A qualitative mapping of the arrival directions of transient acoustic signals. The majority of the signals 
     come from above.  In the lower hemisphere the ANTARES acoustic position system can be identified.  }}
\end{center}
\end{figure}

Figure \ref{fig:sky} shows a qualitative mapping of the arrival directions of transient acoustic signals originating in the surrounding 
of the ANTARES detector. The direction reconstruction was done using the time difference method. The data is from the time period 
29.01.2008 to 05.06.2008\footnote{Most of the data was taken when the detector consisted of 10 lines.}, using a minimum bias trigger which 
is taking unfiltered data for 10 seconds every 30 minutes. Taking this 
data set, the directions of all included signals with an amplitude greater than $8\cdot$RMS (about 200000 signals) were reconstructed. The RMS was 
calculated for a $0.1\, $s time interval around the signal time. A virtual observer residing on the second cluster from the bottom of the $IL07$, about $180$ m 
above the sea bed, looking westward towards the horizon of the storey, is defining the origin of the coordinate system, see Fig. \ref{fig:sky}. 
Thus $90^\circ$ ($-90^\circ$) in longitude ($\phi$) is north (south), while $90^\circ$ ($-90^\circ$) in latitude ($\theta$) is looking upwards (downwards). 
As expected, the majority of the signals is coming from above the detector, including all kinds of transient signals, as e.g. 
produced by dolphins. In the lower hemisphere a few sources can be seen amongst which a clear structure can be identified.  
The Signals between $0^\circ$ and $-120^\circ$ in $\varphi$ and $-30^\circ$ to $-70^\circ$ in $\theta$ originate from the 
ANTARES acoustic positioning system, emitting signals at the bottom of each string.

\section{Source reconstruction}

For a given local cluster such as in the AMADEUS setup, directions are the prime observables of acoustic signals. 
The identified directions then allow for a simple source position reconstruction.
After the successful direction reconstruction, each antenna $i$ that detected a signal 
will point to a direction $\vec{k_i}$ identifying where an event came from. 
Since the cluster positions $\vec{a_i}$ are known, one obtains a line pointing to the event for each cluster,
\begin{equation}
	\vec{d_i}=\vec{a_i}+n_i\vec{k_i} \mbox{ , } n_i \in \mathbb{R}\mbox{.}
\end{equation}
The source position of a signal is theoretically given by the intersection point of the lines $\vec{d_i}$. 
Due to inherent uncertainties in the direction reconstruction, mainly due to the one degree grid, the source location is where the lines are closest to each other.
Localising the point of closest approach for all lines is realized by calculating the square distance from a point 
$\vec{s}$ to the reconstructed lines, 

\begin{equation}
\label{eqn:dist}
	L_i^2(\vec{s})=\left(\vec{s}-(\vec{a_i}+n_i\vec{k_i})\right)^2 \nonumber 
	\mbox{    with } n_i=(\vec{s}-\vec{a_i})\cdot\vec{k_i}\nonumber \mbox{ , }
\end{equation}
and minimising the sum of these distances.

\begin{figure}[ht]
\begin{center}
  \includegraphics[angle=0,width=0.49\textwidth]{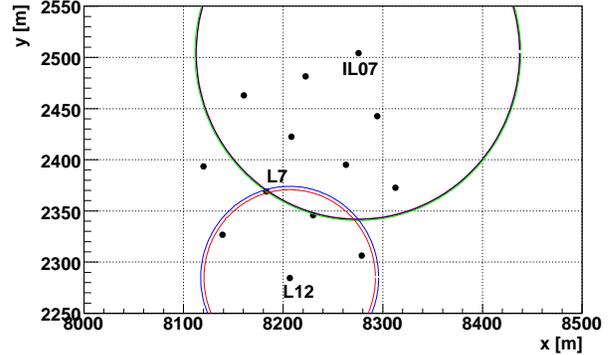}
  \caption {\label{fig:source} {Layout of ANTARES on the sea floor in UTM coordinates.  Each point indicates the position of a line.  The positions of the 
     Instrumentation Line (IL07), Line 12 (L12) and Line 7 (L7) are labeled. 
  }}
\end{center}
\end{figure}

Since the cluster positioning and orientation was not yet implemented, a full source position reconstruction was not possible at the time of this analysis. 
However, as the tilts of the local clusters, induced by the sea current, are small, in the order of one degree, it is possible to reconstruct the $\theta$ angle with 
that uncertainty. The circles in Fig.\ \ref{fig:source} represent the possible solutions of a source given by the reconstructed $\theta$ angle and the distance 
of each storey to the sea floor. The signal origin can clearly be identified as the acoustic positioning emitter of the 7th line (L7), as 
can also be deduced from the signal properties.

\section{Summary}
Techniques for source reconstruction using localized clusters have been presented. Source location is done by using the local 
clusters for direction reconstruction.  
For direction reconstruction two methods were investigated.  One is a beam-forming algorithm only requiring the hydrophone positions, within a local cluster, 
and synchronized data. The other method additionally requires a signal identification and uses time differences.
Although both methods work well, using time differences proved to be much faster.
A technique for source reconstruction, based on the directional information of the local clusters, was explained and its potential demonstrated.  
\\This study was supported by the German government through BMBF grant 05CN5WE1/7.


\begin{thebibliography}{00}
\bibitem{arena06}{K. Graf et al., Integration of Acoustic Neutrino Detection Methods into ANTARES, J. Phys, 2007, Conf. Ser. 81 012012}
\bibitem{antares_proposal}{E. Aslanides et al., A Deep Sea Telescope for High Energy Neutrinos, 1999; arXiv:astro-ph/9907432 }
\end{thebibliography}
\end{document}